\documentclass[onecolumn,useAMS,usenatbib]{mn2e}

\usepackage{graphicx}
\usepackage{subfigure}
\usepackage{xcolor}
\usepackage{mathtext,bbm,amsmath,amsfonts,amssymb,indentfirst,syntonly,graphicx}
\usepackage{mathtools}
\usepackage{slashbox}
\usepackage[english]{babel}
\usepackage{calc}
\usepackage{tikz}

\def\bc{\begin{center}}
\def\ec{\end{center}}
\def\be{\begin{eqnarray}}
\def\ee{\end{eqnarray}}

\title[Distance calibration of GRBs and constrain on the $\Lambda$CDM]{Model-independent distance calibration of high-redshift gamma-ray bursts and constrain on the $\Lambda$CDM model}
\author[H.-N. Lin, X. Li and Z. Chang]
        {Hai-Nan Lin$^{1}$\thanks{e-mail: linhn@ihep.ac.cn.},
        Xin Li$^{1,3}$\thanks{e-mail: lixin1981@cqu.edu.cn.},
        Zhe Chang$^{2}$\thanks{e-mail: changz@ihep.ac.cn.}\\
$^{1}$Department of Physics, Chongqing University, Chongqing 401331, China\\
$^{2}$Institute of High Energy Physics, Chinese Academy of Sciences, Beijing 100049, China\\
$^{3}$State Key Laboratory Theoretical Physics, Institute of Theoretical Physics, Chinese Academy of Sciences, Beijing 100190, China}
\begin{document}

\date{Accepted xxxx; Received xxxx; in original form xxxx}

\pagerange{\pageref{firstpage}--\pageref{lastpage}} \pubyear{2015}

\maketitle

\label{firstpage}

\begin{abstract}
Gamma-ray bursts (GRBs) are luminous enough to be detectable up to redshift $z\sim 10$. They are often proposed as complementary tools to type-Ia supernovae (SNe Ia) in tracing the Hubble diagram of the Universe. The distance calibrations of GRBs usually make use one or some of the empirical luminosity correlations, such as $\tau_{\rm lag}-L$, $V-L$, $E_p-L$, $E_p-E_{\gamma}$, $\tau_{\rm RT}-L$ and $E_p-E_{\rm iso}$ relations. These calibrating methods are based on the underling assumption that the empirical luminosity correlations are universal over all redshift range. In this paper, we test the possible redshift dependence of six luminosity correlations by dividing GRBs into low-$z$ and high-$z$ classes according to their redshift smaller or larger than 1.4. It is shown that the $E_p-E_{\gamma}$ relation for low-$z$ GRBs is consistent with that for high-$z$ GRBs within $1\sigma$ uncertainty. The intrinsic scatter of $V-L$ relation is too larger to make a convincing conclusion. For the rest four correlations, however, low-$z$ GRBs differ from high-$z$ GRBs at more than $3\sigma$ confidence level. As such, we calibrate GRBs using the $E_p-E_{\gamma}$ relation in a model-independent way. The constraint of high-$z$ GRBs on the $\Lambda$CDM model gives $\Omega_M=0.302\pm 0.142(1\sigma)$, well consistent with the Planck 2015 results.
\end{abstract}


\section{Introduction}\label{sec:introduction}

The consistent luminosity of type-Ia supernovae (SNe Ia) makes them to be standard candles in probing the expansion history of the Universe \citep{Riess:1998,Perlmutter:1999}. However, the redshift of SNe Ia is usually less than 1.4. Gamma-ray bursts (GRBs), as the most energetic explosions in the Universe, are bright enough to be detectable up to redshift $z\sim 10$ \citep{Salvaterra:2009,Tanvir:2009,Cucchiara:2011, Tanvir:2013}. Therefore, they are often proposed as complementary tools to SNe Ia in tracing the Hubble diagram of the high-redshift Universe. Actually, GRBs have already been widely used, either alone or in combination with other data such as SNe Ia, to constrain the cosmological parameters \citep{Schaefer:2003,Bloom:2003,Xu:2005,Firmani:2005,Liang:2005,Firmani:2006a,Schaefer:2007,Liang:2008a,Liang:2008b,Wei:2009,Wei:2010,Demianski:2011, Wang:2011,Capozziello:2012,Amati:2013sca,Wei:2013,Velten:2013,Cai:2013,Breton:2013, Chang:2014,Cano:2014,Cuzinatto:2014,Wang:2014,Wang:2015,Li:2015}.
Unfortunately, since the explosion mechanism of GRBs is still not clearly known, the distance calibration of GRBs is not as easy as that of SNe Ia.

Many methods have been proposed to calibrate GRBs \citep{Dai:2004tq,Ghirlanda:2004a,Liang:2005,Firmani:2005,Schaefer:2007,Liang:2008a,Liang:2008b,Wei:2009,Wei:2010,Liu:2014}. Most calibrating methods rely on the empirical luminosity correlations found in long GRBs. At least six luminosity correlations can be used in the calibration. \citet{Norris:2000} found a correlation between spectrum lag and isotropic peak luminosity ($\tau_{\rm lag}-L$ relation). \citet{Fenimore:2000} found a correlation between time variability and isotropic peak luminosity ($V-L$ relation). \citet{Amati:2002} found a tight correlation between the peak energy of $\nu F_{\nu}$ spectrum and isotropic equivalent energy ($E_p-E_{\rm iso}$ relation). \citet{Ghirlanda:2004b} found a similar correlation between peak energy and collimation-corrected energy ($E_p-E_{\gamma}$ relation). \citet{Yonetoku:2004} found a correlation between peak energy and isotropic peak luminosity ($E_p-L$ relation). \citet{Schaefer:2007} found a correlation between minimum rise time of light curve and isotropic peak luminosity ($\tau_{\rm RT}-L$ relation).

All of the calibrating methods based on the empirical luminosity correlations have an underlying assumption, that is, the luminosity correlations do not evolve with redshift. If the luminosity correlations is not universal over the whole redshift range, these calibrating methods will fail. In fact, the possible redshift dependence of luminosity correlations has already been tested by some authors. \citet{Basilakos:2008} investigated the above six empirical luminosity correlations in four redshift bins, and showed that the slopes of all six correlations differs between redshift bins, although the intercepts do not vary significantly. Since the GRB sample is not large enough in each bin, the statistical uncertainty is large. Therefore, they concluded that no statistically significant evidence for the redshift evolution of the luminosity correlations was found. With the updated data, \citet{Wang:2011} got a similar conclusion. However, \citet{Li:2007} investigated the Amati relation in four redshift bins and showed that the slope and intercept varies with redshift systematically and significantly. Recently, \citet{Lin:2015} divided GRBs into two redshift bins, and found that the Amati relation (especially the slope parameter) of low-$z$ GRBs differs from that of high-$z$ GRBs at more than $3\sigma$ confidence level. \citet{Dainotti:2013} investigated the slope evolution of GRB correlations and showed that correlation slope that differs from the intrinsic one may overestimate or underestimate the cosmological parameters.

In this paper, we recheck the possible redshift dependence of six luminosity correlations. We divide GRBs into low-$z$ and high-$z$ classes according to their redshift smaller or larger than 1.4, and test the luminosity correlations for low-$z$ and high-$z$ GRBs, respectively. The main difference between our work and \citet{Wang:2011}'s is that we just divide GRBs into two redshift bins, so that the number of GRBs in each bin is large enough to do statistical analysis. We choose $z=1.4$ as the threshold because the redshift of SNe Ia is usually smaller than 1.4, and the Universe below this redshift has already been tightly constrained. We find that, among the six luminosity correlations, only the $E_p-E_{\gamma}$ relation is consistent between low-$z$ and high-$z$ GRBs within $1\sigma$ uncertainty. As such, we can calibrate GRBs through the $E_p-E_{\gamma}$ relation using the Pad\'{e} approximation proposed by \citet{Liu:2014}, and the Hubble diagram of GRBs can be constructed.

The rest of the paper is arranged as follows: In section \ref{sec:correlations}, we test the redshift dependence of six luminosity correlations. In section \ref{sec:calibration}, we calibrate the distance of high-$z$ GRBs using the $E_p-E_{\gamma}$ relation, and then use them to constrain the $\Lambda$CDM model. Finally, a short summary is given in section \ref{sec:conclusions}.

\section{Testing the redshift dependence of luminosity correlations}\label{sec:correlations}

All the six luminosity correlations mentioned above have the exponential form $R=AQ^b$, which can be linearized by taking the logarithm, i.e.,
\begin{equation}
  y=a+bx \qquad (y\equiv\log R,~x\equiv\log Q,~a\equiv\log A),
\end{equation}
where ``\,$\log$" represents the logarithm of base 10. For the sake of clarity, we write the six luminosity correlations explicitly here:
\begin{align}\label{eq:correlation}
\log\frac{L}{{\rm erg~s}^{-1}}&=a_1+b_1\log\frac{\tau_{{\rm lag},i}}{0.1~{\rm s}},\\
\log\frac{L}{{\rm erg~s}^{-1}}&=a_2+b_2\log\frac{V_i}{0.02},\\
\log\frac{L}{{\rm erg~s}^{-1}}&=a_3+b_3\log\frac{E_{p,i}}{300~{\rm keV}},\\ \label{eq:Ep-Egamma}
\log\frac{E_{\gamma}}{{\rm erg}}&=a_4+b_4\log\frac{E_{p,i}}{300~{\rm keV}},\\
\log\frac{L}{{\rm erg~s}^{-1}}&=a_5+b_5\log\frac{\tau_{{\rm RT},i}}{0.1~{\rm s}},\\
\log\frac{E_{\rm iso}}{{\rm erg}}&=a_6+b_6\log\frac{E_{p,i}}{300~{\rm keV}},
\end{align}
where quantities with a subscript ``\,$i$\," represent the quantities in the comoving frame, which can be transformed to the observer frame by $\tau_{{\rm lag},i}=\tau_{\rm lag}(1+z)^{-1}$, $\tau_{{\rm RT},i}=\tau_{\rm RT}(1+z)^{-1}$, $V_i=V(1+z)$, and $E_{p,i}=E_p(1+z)$.

The isotropic peak luminosity $L$ can be calculated from the bolometric peak flux $P_{\rm bolo}$ as \citep{Schaefer:2007}
\begin{equation}\label{eq:iso_luminosity}
  L=4\pi d_L^2P_{\rm bolo},
\end{equation}
where $d_L$ is the luminosity distance. The bolometric peak flux $P_{\rm bolo}$ is calculated from the observed peak photon flux in the rest frame $1-10,000$ keV energy band by assuming the Band spectrum \citep{Band:1993}. The luminosity distance depends on a specific cosmological model. In the concordance $\Lambda$CDM model, it is given as
\begin{equation}\label{lumi_distance}
  d_L(z)=(1+z)\frac{c}{H_0}\int_0^{z} \frac{dz}{\sqrt{\Omega_M(1+z)^3+(1-\Omega_M)}},
\end{equation}
where $\Omega_M$ is the mater density, $H_0$ is the Hubble constant, and $c$ is the light speed. Here we take $\Omega_M=0.280$ and $H_0=70.0~\rm{km}~\rm{s}^{-1}~\rm{Mpc}^{-1}$ from fitting to the Union2.1 dataset \citep{Lin:2015}. The uncertainty of $L$  propagates from the uncertainty of $P_{\rm bolo}$, while that from $d_L$ is absorbed into the intrinsic scatter. The isotropic equivalent energy $E_{\rm iso}$ can be calculated from the bolometric fluence $S_{\rm bolo}$ as \citep{Schaefer:2007}
\begin{equation}\label{eq:iso_energy}
  E_{\rm iso}=4\pi d_L^2S_{\rm bolo}(1+z)^{-1}.
\end{equation}
Similar to the bolometric peak flux, the bolometric fluence $S_{\rm bolo}$ also corresponds to the rest frame $1-10,000$ keV energy band. For $E_{\rm iso}$, we also only consider the error propagation from $S_{\rm bolo}$. The collimation-corrected energy, $E_{\gamma}$, is the isotropic equivalent energy multiplied by a beaming factor $F_{\rm beam}\equiv1-\cos\theta_{\rm jet}$, where $\theta_{\rm jet}$ is the jet opening angle, i.e,
\begin{equation}\label{eq:collimation_energy}
  E_{\gamma}\equiv E_{\rm iso}F_{\rm beam}=4\pi d_L^2S_{\rm bolo}F_{\rm beam}(1+z)^{-1}.
\end{equation}
The uncertainty of $E_{\gamma}$ propagates from the uncertainties of both $S_{\rm bolo}$ and $F_{\rm beam}$. The error propagation from $Q$ to $\log Q$ is given as
\begin{equation}
  \sigma_{\log Q}=\frac{1}{\ln 10}\frac{\sigma_Q}{Q},
\end{equation}
where ``\,$\ln$" represents the natural logarithm. If $Q$ has nonsymmetric error, we symmetrize it by taking the average, i.e., $\sigma_Q=(\sigma_Q^+ + \sigma_Q^-)/2$.

To test the possible redshift dependence of luminosity correlations, we analyze the GRB sample taken from \citet{Wang:2011}. This sample consists of 116 long GRBs in the redshift range $z\in [0.17,8.2]$. This dataset is a collection of GRBs with well-measured spectra properties from various instruments, such as BATSE, Konus, Swift, etc.. We divide GRBs into two subsamples according to their redshift smaller or larger than 1.4, and call them low-$z$ and high-$z$ subsamples, respectively. We choose $z=1.4$ as the threshold because the redshift of SNe Ia is usually smaller than 1.4. The Universe below this redshift has already been well studied using SNe Ia \citep{Amanullah:2010,Suzuki:2012,Betoule:2014}. Low-$z$ and high-$z$ subsamples consist of 50 and 66 GRBs, respectively. We fit each luminosity correlation to the two subsamples separately. Since the plot of each correlation in the $xy$ plane show large error bars in both the horizontal and vertical axes, and intrinsic scatter dominates over the measurement error, the ordinary least-$\chi^2$ method does not work well. We apply the fitting method presented in \citet{DAgostini:2005}. The best-fit parameters ($a,b,\sigma_{\rm int}$) can be derived by maximizing the D'Agostini's likelihood,
\begin{equation}\label{eq:likelihood}
  \mathcal{L}_D(\sigma_{\rm int},a,b)\propto\prod_i\frac{1}{\sqrt{\sigma_{\rm int}^2+\sigma_{y_i}^2+b^2\sigma_{x_i}^2}}
  \times \exp\left[-\frac{(y_i-a-bx_i)^2}{2(\sigma_{\rm int}^2+\sigma_{y_i}^2+b^2\sigma_{x_i}^2)}\right],
\end{equation}
where the intrinsic scatter $\sigma_{\rm int}$ represents any other unknown errors except for the measurement error. Equivalently, we can minimizing the $\chi^2$,
\begin{equation}\label{eq:loglikelihood}
  \chi_D^2(\sigma_{\rm int},a,b)=\sum_i\ln(\sigma_{\rm int}^2+\sigma_{y_i}^2+b^2\sigma_{x_i}^2) + \sum_i\frac{(y_i-a-bx_i)^2}{\sigma_{\rm int}^2+\sigma_{y_i}^2+b^2\sigma_{x_i}^2}.
\end{equation}

\begin{figure}
  \centering
 \includegraphics[width=16 cm, height=18.8 cm]{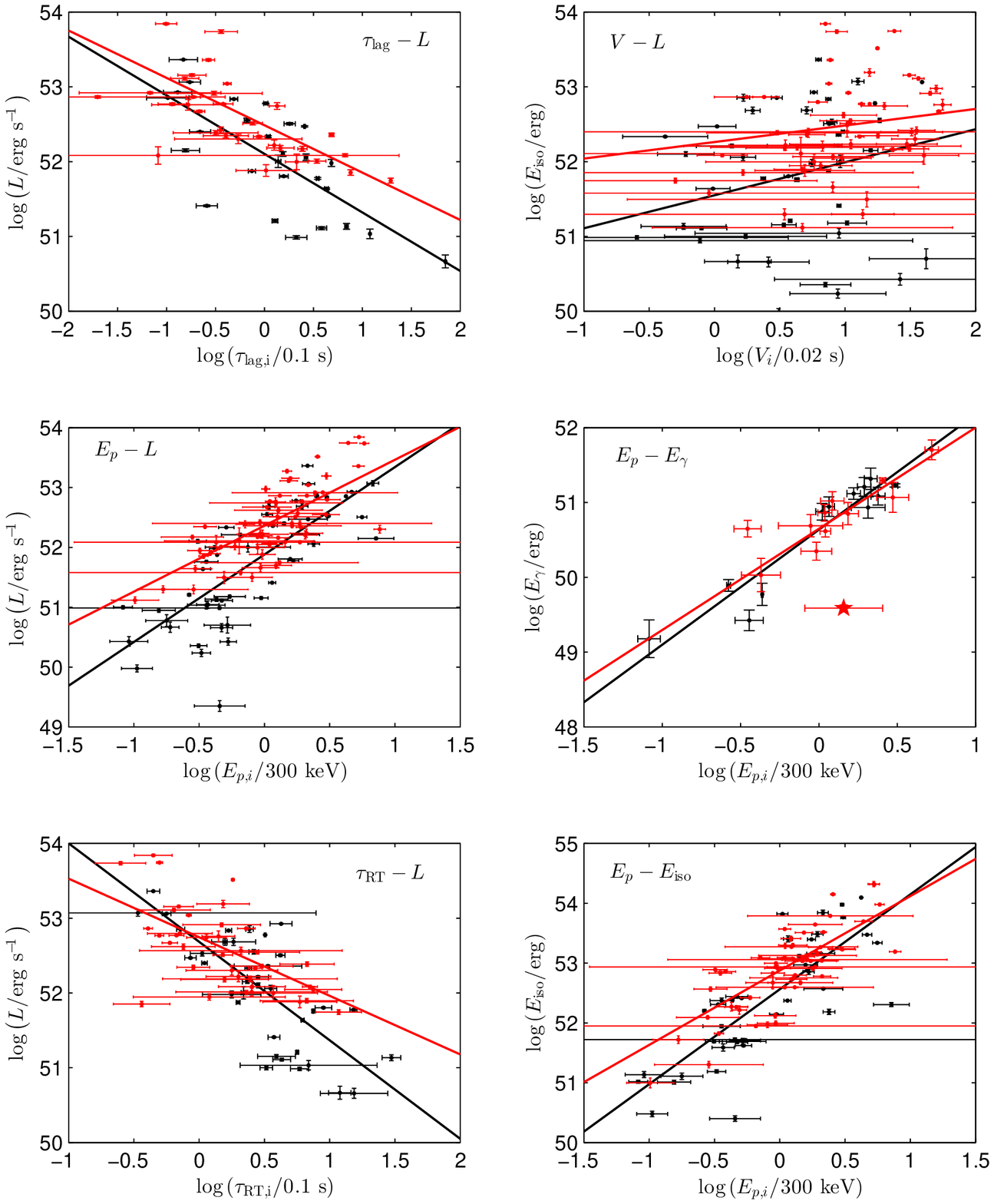}
 \caption{\small{The luminosity correlations for low-$z$ (black) and high-$z$ (red) GRBs. Error bars represent the $1\sigma$ uncertainties. The lines are the best-fit results, which are derived from maximizing the D'Agostini's likelihood.}}\label{fig:correlation}
\end{figure}

\begin{figure}
  \centering
 \includegraphics[width=16 cm, height=18.8 cm]{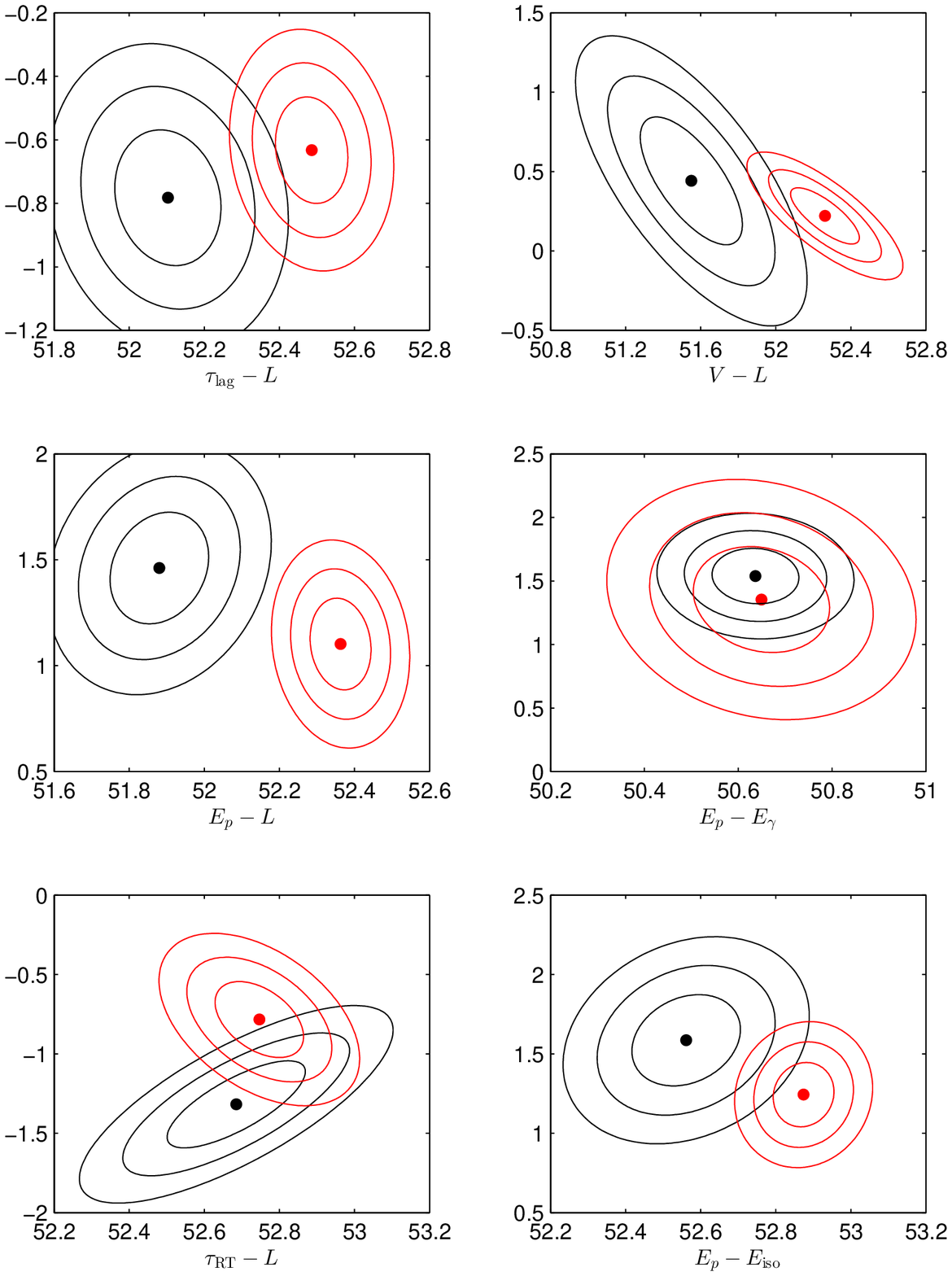}
 \caption{\small{The $1\sigma$, $2\sigma$ and $3\sigma$ contours in the $(a,b)$ plane for low-$z$ (black curves) and high-$z$ (red curves) GRBs derived from the D'Agostini's likelihood. The central values are denoted by dots.}}\label{fig:contours}
\end{figure}

We use the publicly available Matlab package FMINUIT\footnote{http://www.fis.unipr.it/$\sim$giuseppe.allodi/Fminuit/Fminuit\b{\,\,}intro.html} to derive the best-fit parameters and their uncertainties. The results are listed in the fourth to sixth columns in Table \ref{tab:parameters}. This table gives the mean values of the best-fit parameters and their $1\sigma$ uncertainties.
\begin{table}
\centering
\caption{\small{The intrinsic scatters ($\sigma_{\rm int}$), intercepts ($a$) and slopes ($b$) of six luminosity correlations for low-$z$ and high-$z$ GRBs, derived from maximizing the D'Agostini's likelihood. The quoted errors are of $1\sigma$. $N$ is the number of GRBs available in the fitting.}}\label{tab:parameters}
\begin{tabular}{ll|cccc}
  \hline\hline
   (1) & (2) & (3) & (4) & (5) & (6) \\
   correlation & subsample  & $N$ & $\sigma_{\rm int}$ & $a$ & $b$ \\
  \hline
  $\tau_{\rm lag}-L$:   & low-$z$  & 27 & $0.475\pm 0.066$ & $52.103\pm 0.093$ & $-0.783\pm 0.141$ \\
                        & high-$z$ & 32 & $0.321\pm 0.049$ & $52.486\pm 0.064$ & $-0.633\pm 0.111$ \\
  \hline
  $V-L$:                & low-$z$  & 47 & $0.875\pm 0.101$ & $51.550\pm 0.179$ & $0.441\pm 0.266$ \\
                        & high-$z$ & 57 & $0.546\pm 0.057$ & $52.262\pm 0.121$ & $0.221\pm 0.117$ \\
  \hline
  $E_p-L$:              & low-$z$  & 50 & $0.577\pm 0.063$ & $51.880\pm 0.087$ & $1.461\pm 0.174$ \\
                        & high-$z$ & 66 & $0.386\pm 0.039$ & $52.363\pm 0.054$ & $1.102\pm 0.143$ \\
  \hline
  $E_p-E_{\gamma}$:     & low-$z$  & 12 & $0.159\pm 0.059$ & $50.637\pm 0.061$ & $1.539\pm 0.144$ \\
                        & high-$z$ & 12 & $0.261\pm 0.105$ & $50.649\pm 0.096$ & $1.354\pm 0.275$ \\
  \hline
  $\tau_{\rm RT}-L$:    & low-$z$  & 39 & $0.470\pm 0.058$ & $52.685\pm 0.122$ & $-1.318\pm 0.181$ \\
                        & high-$z$ & 40 & $0.395\pm 0.052$ & $52.747\pm 0.078$ & $-0.784\pm 0.158$ \\
  \hline
  $E_p-E_{\rm iso}$:    & low-$z$  & 40 & $0.561\pm 0.069$ & $52.561\pm 0.095$ & $1.586\pm 0.190$ \\
                        & high-$z$ & 61 & $0.365\pm 0.040$ & $52.874\pm 0.053$ & $1.243\pm 0.134$ \\
  \hline
\end{tabular}
\end{table}
Note that not all GRBs are available in the analysis of each luminosity correlation. For example, GRBs without measurement of the jet opening angle is unavailable in the $E_p-E_{\gamma}$ analysis, while GRBs having no spectrum lag measurement are invalid in the $\tau_{\rm lag}-L$ analysis. For this reason, we also list the number of available GRBs in each fitting in the third column of Table \ref{tab:parameters}. All the six luminosity correlations are plotted in Figure \ref{fig:correlation} in logarithmic coordinates. Low-$z$ and high-$z$ GRBs are denoted by black and red dots, respectively. The error bars represent $1\sigma$ uncertainties. Since the Swift/BAT instrument is only sensitive in a narrow energy band ($\sim 15-150$ keV), the uncertainties of peak energy of some GRBs are extremely large. The lines stand for the best-fit results (black line for low-$z$ GRBs and red line for high-$z$ GRBs).
Besides, we also plot the $1\sigma$, $2\sigma$ and $3\sigma$ contours in the $(a,b)$ plane for low-$z$ (black curves) and high-$z$ (red curves) GRBs in Figure \ref{fig:contours}. The best-fit central values are denoted by dots.

From Table \ref{tab:parameters} and Figure \ref{fig:correlation}, we can see that among six luminosity correlations, the $V-L$ relation has the largest intrinsic scatter, while the $E_p-E_{\gamma}$ relation has the smallest intrinsic scatter. The intrinsic scatter of the $V-L$ relation is so large that it is unreasonable to fit it with a line. For all the six luminosity correlations, high-$z$ GRBs have larger intercept, but smaller absolute slope than low-$z$ GRBs, although the difference of intercepts between low-$z$ and high-$z$ GRBs is not as significant as that of slopes. The slope difference of the $\tau_{\rm RT}-L$ relation is especially evident. This can be seen more clearly from the contour plots in the ($a,b$) plane in Figure \ref{fig:contours}. The $E_p-E_{\gamma}$ relation of low-$z$ GRBs is consistent with that of high-$z$ GRBs within $1\sigma$ uncertainty. However, for the rest five luminosity correlations, low-$z$ GRBs differ from high-$z$ GRBs at more than $3\sigma$ confidence level. Especially, there is no overlap between the $3\sigma$ contours of two subsamples for the $E_p-L$ relation. As for the Amati relation, we recover the results of \citet{Lin:2015}.

The results above are derived using D'Agostini's likelihood. Since the observed data points have significant errors on both the $x$-axis and $y$-axis, there is no unique method to determine the best-fit parameters. \citet{Reichart:2001} has constructed a likelihood which is slightly different from D'Agostini's one. To test whether the above results depend on the choice of a specific best-fit method, we also do a similar calculation using Reichart's likelihood. The Reichart's likelihood is written as \citep{Reichart:2001}
\begin{equation}\label{eq:likelihood2}
  \mathcal{L}_R(\sigma_x,\sigma_y,a,b)\propto\prod_i\frac{\sqrt{1+b^2}}{\sqrt{\sigma_y^2+\sigma_{y_i}^2+b^2(\sigma_x^2+\sigma_{x_i}^2)}}
  \times \exp\left[-\frac{(y_i-a-bx_i)^2}{2[\sigma_y^2+\sigma_{y_i}^2+b^2(\sigma_x^2+\sigma_{x_i}^2)]}\right],
\end{equation}
where $\sigma_x$ and $\sigma_y$ are the intrinsic scatters along the $x$-axis and $y$-axis, respectively. The corresponding $\chi^2$ is given as
\begin{equation}\label{eq:loglikelihood2}
  \chi_R^2(\sigma_x,\sigma_y,a,b)=\sum_i\ln[\sigma_y^2+\sigma_{y_i}^2+b^2(\sigma_x^2+\sigma_{x_i}^2)] + \sum_i\frac{(y_i-a-bx_i)^2}{\sigma_y^2 +\sigma_{y_i}^2+b^2(\sigma_x^2+\sigma_{x_i}^2)}-N\ln(1+b^2),
\end{equation}
where $N$ is the number of data points. The best-fit parameters are the one which can minimize the right-hand-side of Eq.(\ref{eq:loglikelihood2}).

\begin{figure}
  \centering
 \includegraphics[width=16 cm, height=18.8 cm]{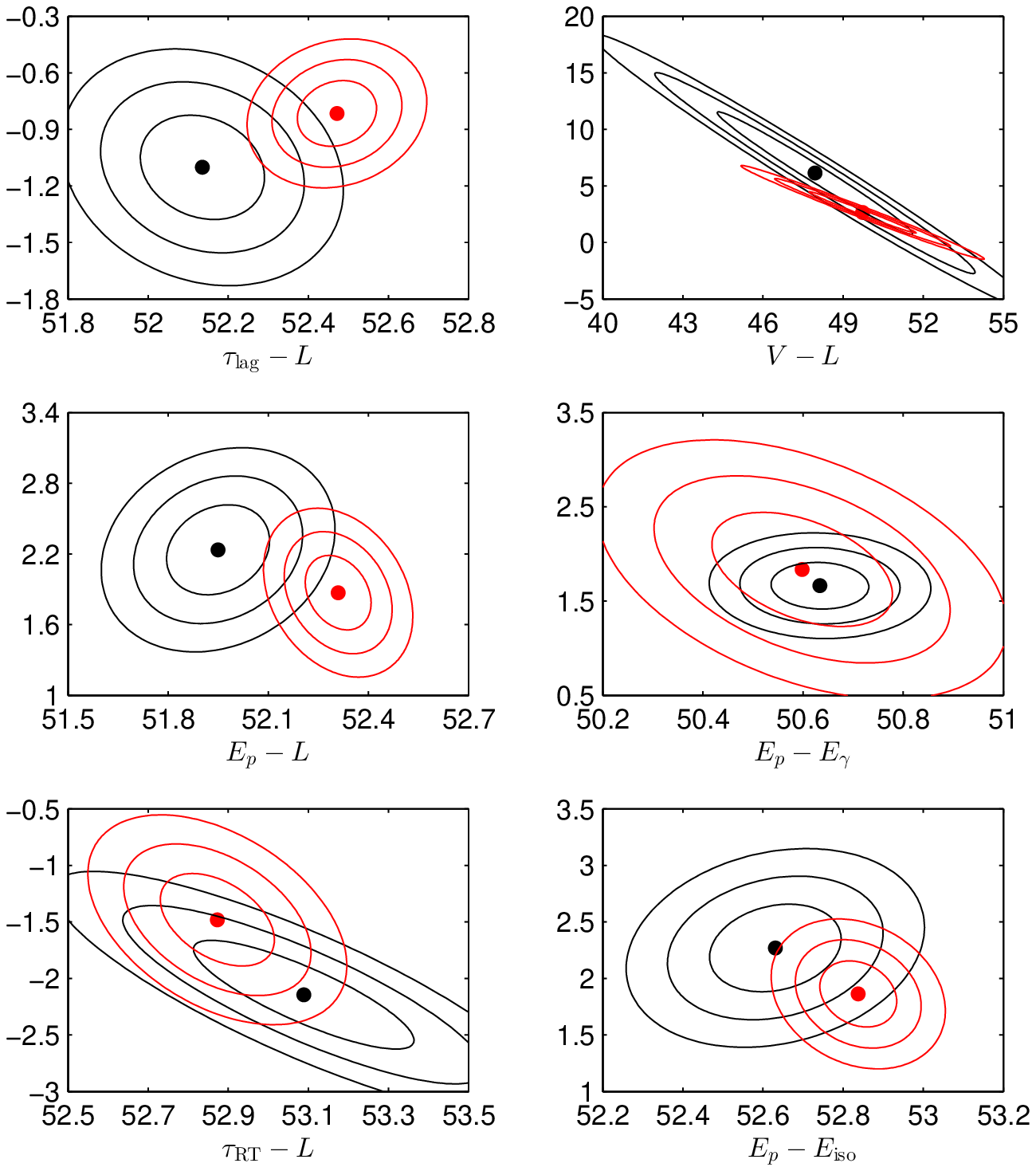}
 \caption{\small{The $1\sigma$, $2\sigma$ and $3\sigma$ contours in the $(a,b)$ plane for low-$z$ (black curves) and high-$z$ (red curves) GRBs derived from the Reichart's likelihood. The central values are denoted by dots.}}\label{fig:contours2}
\end{figure}

The best-fit parameters and their $1\sigma$ uncertainties are listed in Table \ref{tab:parameters2}.
\begin{table}
\centering
\caption{\small{The intrinsic scatters along the $x$-axis ($\sigma_x$) and $y$-axis ($\sigma_y$), intercepts ($a$) and slopes ($b$) of six luminosity correlations for low-$z$ and high-$z$ GRBs, derived from maximizing the Reichart's likelihood. The quoted errors are of $1\sigma$. $N$ is the number of GRBs available in the fitting. The last column gives the total intrinsic scatter $\sigma_{\rm int}\equiv(\sigma_y^2+b^2\sigma_x^2)^{1/2}$.}}\label{tab:parameters2}
\begin{tabular}{ll|cccccc}
  \hline\hline
   (1) & (2) & (3) & (4) & (5) & (6) & (7) & (8)\\
   correlation & subsample  & $N$ & $\sigma_{x}$ & $\sigma_{y}$ & $a$ & $b$ & $\sigma_{\rm int}$ \\
  \hline
  $\tau_{\rm lag}-L$:   & low-$z$  & 27 & $0.271 \pm 1.757 $ &  $0.417 \pm 1.385 $	&	$52.136 \pm 0.102 $	&	$-1.101 \pm 0.183 $	&	$0.513 \pm 1.593 $ \\
                        & high-$z$ & 32 & $0.236 \pm 1.390 $ &	$0.255 \pm 0.855 $	&	$52.471 \pm 0.065 $	&	$-0.815 \pm 0.115 $	&	$0.319 \pm 0.966 $ \\
  \hline
  $V-L$:                & low-$z$  & 47 & $0.395 \pm 0.055 $ &	$0.010 \pm 29.430 $	&	$47.945 \pm 2.413 $	&   $6.135 \pm 3.571 $	&	$2.425 \pm 1.457 $ \\
                        & high-$z$ & 57 & $0.390 \pm 0.754 $ &	$0.236 \pm 8.803 $	&	$49.730 \pm 1.324 $	&	$2.661 \pm 1.206 $	&	$1.064 \pm 2.804 $ \\
  \hline
  $E_p-L$:              & low-$z$  & 50 & $0.042 \pm 3.396 $ &	$0.666 \pm 1.053 $	&	$51.950 \pm 0.102 $	&	$2.235 \pm 0.252 $	&	$0.672 \pm 1.479 $ \\
                        & high-$z$ & 66 & $0.091 \pm 1.036 $ &	$0.396 \pm 0.829 $	&	$52.309 \pm 0.065 $	&	$1.871 \pm 0.208 $	&	$0.431 \pm 1.078 $ \\
  \hline
  $E_p-E_{\gamma}$:     & low-$z$  & 12 & $0.047 \pm 1.121 $ &	$0.152 \pm 0.951 $	&	$50.634 \pm 0.064 $	&	$1.663 \pm 0.163 $	&	$0.171 \pm 1.198 $ \\
                        & high-$z$ & 12 & $0.159 \pm 0.931 $ &	$0.073 \pm 6.867 $	&	$50.598 \pm 0.119 $	&	$1.836 \pm 0.399 $	&	$0.300 \pm 2.347 $ \\
  \hline
  $\tau_{\rm RT}-L$:    & low-$z$  & 39 & $0.073 \pm 1.438 $ &	$0.551 \pm 0.880 $	&	$53.088 \pm 0.182 $	&	$-2.144 \pm 0.317 $	&	$0.573 \pm 1.196 $ \\
                        & high-$z$ & 40 & $0.222 \pm 0.778 $ &	$0.293 \pm 1.291 $	&	$52.873 \pm 0.094 $	&	$-1.481 \pm 0.270 $	&	$0.440 \pm 1.217 $ \\
  \hline
  $E_p-E_{\rm iso}$:    & low-$z$  & 40 & $0.225 \pm 0.658 $ &	$0.373 \pm 2.042 $	&	$52.631 \pm 0.108 $	&	$2.270 \pm 0.255 $	&	$0.632 \pm 1.705 $ \\
                        & high-$z$ & 61 & $0.160 \pm 0.474 $ &	$0.275 \pm 0.955 $	&	$52.838 \pm 0.063 $	&	$1.863 \pm 0.193 $	&	$0.405 \pm 0.917 $ \\
  \hline
\end{tabular}
\end{table}
The last column gives the ``equivalent" total intrinsic scatter, which is calculated from $\sigma_{\rm int}\equiv(\sigma_y^2+b^2\sigma_x^2)^{1/2}$. Comparing to Table \ref{tab:parameters}, we can see from Table \ref{tab:parameters2} that the $E_p-E_{\gamma}$ relation has the smallest (while the $V-L$ relation has the largest) intrinsic scatters, although the uncertainties of intrinsic scatters in Table \ref{tab:parameters2} are much larger. Using Reichart's likelihood, the parameters (especially the intrinsic scatter) cannot be well constrained. Reichart's likelihood leads to larger absolute slope parameters compared to D'Agostini's likelihood. Figure \ref{fig:contours2} is the contour plot in the $(a,b)$ plane.
We can see an important common feature between the results derived from two different likelihoods: only for the $E_p-E_{\gamma}$ relation, low-$z$ subsample is consistent with high-$z$ subsample within $1\sigma$ uncertainty. For the $\tau_{\rm lag}-L$ and $E_p-L$ relations, low-$z$ subsample still differs from high-$z$ subsample at more than $3\sigma$ confidence level. As for the $\tau_{\rm RT}-L$ and $E_p-E_{\rm iso}$ relations, low-$z$ subsample differs from high-$z$ subsample at more than $2\sigma$ confidence level. The relatively lower significance is due to the larger uncertainties of the best-fit parameters. The uncertainties of slope parameters of $V-L$ relation derived from Reichart's likelihood are extremely large. In a word, only $E_p-E_{\gamma}$ relation shows no significant evidence for the redshift evolution. This conclusion does not depend on the choice of the best-fit methods.

Reichart's likelihood differs from D'Agostini's one by an extra factor $(1+b^2)^{1/2}$. Otherwise, these two likelihoods are identical (if we set $\sigma_{\rm int}^2\equiv \sigma_y^2+b^2\sigma_x^2$). \citet{DAgostini:2005} pointed out that Reichart's likelihood has a problem: $m^2$ cannot be added tout court to 1, since $m^2$ is in general dimensional (although in our case it is dimensionless). The factor $(1+b^2)^{1/2}$ has the net effect of overestimating $m$. This is one reason why Reichart's likelihood leads to a larger slope parameters relative to D'Agostini's likelihood. Therefore, we use the results of D'Agostini's likelihood when calibrating the distance of GRBs in the next section.

\section{Distance calibration and cosmological implications}\label{sec:calibration}

As we have shown that the $E_p-E_{\gamma}$ relation does not significantly evolve with redshift, we can use it to calibrate GRBs. To avoid the circularity problem, the Pad\'{e} method proposed by \citet{Liu:2014} is applied. The main calibrating procedures are as follows: Firstly, derive the distance-redshift relation of SNe Ia (here we use the Union2.1 \citep{Suzuki:2012} dataset) using the Pad\'{e} approximation of order (3,2), i.e.,
\begin{equation}\label{eq:pade}
  \mu(z)=\frac{\alpha_0+\alpha_1z+\alpha_2z^2+\alpha_3z^3}{1+\beta_1z+\beta_2z^2},
\end{equation}
where the coefficients $(\alpha_0, \alpha_1, \alpha_2, \alpha_3, \beta_1, \beta_2)$ and the corresponding covariance matrix are derived by fitting Eq.(\ref{eq:pade}) to the Union2.1 dataset (see \citet{Liu:2014} for details). Assuming that the low-$z$ GRBs trace the same Hubble diagram to SNe Ia, we can calculate the distance moduli of low-$z$ GRBs directly from Eq.(\ref{eq:pade}). The uncertainty of $\mu$ propagates from the uncertainties of the coefficients $(\alpha_i, \beta_i)$. Then the luminosity distance of low-$z$ GRBs can be obtained using the relation
\begin{equation}\label{eq:dis_modulus}
\mu(z)=5\log\frac{d_L(z)}{\rm{Mpc}}+25.
\end{equation}
As $d_L$ is known, the collimation-corrected energy can be further calculated from Eq.(\ref{eq:collimation_energy}). Note that there are only 12 low-$z$ GRBs and 12 high-$z$ GRBs available since the others have no measurement of jet opening angle. Then we fit the $E_p-E_{\gamma}$ relation (i.e., Eq.(\ref{eq:Ep-Egamma})) to the 12 low-$z$ GRBs, which gives the best-fit parameters
\begin{equation}
  \sigma_{\rm int}=0.161\pm 0.059,~~a=50.632\pm 0.062,~~b=1.537\pm 0.145.
\end{equation}
By directly extrapolating the $E_p-E_{\gamma}$ relation to high-$z$ GRBs, we can inversely obtain the collimation-corrected energy for 12 high-$z$ GRBs from Eq.(\ref{eq:Ep-Egamma}). Finally, calculate the luminosity distance of high-$z$ GRBs from Eq.(\ref{eq:collimation_energy}), and then the distance moduli from Eq.(\ref{eq:dis_modulus}). The uncertainty of distance moduli propagates from the uncertainties of $E_{\gamma}$, $S_{\rm bolo}$ and $F_{\rm beam}$, i.e. \citep{Schaefer:2007},
\begin{equation}
  \sigma_{\mu}^2=\left(\frac{5}{2\ln 10}\right)^2\left[(\ln 10)^2\sigma_{\log E_{\gamma}}^2 + \frac{\sigma_{S_{\rm bolo}}^2}{S_{\rm bolo}^2} + \frac{\sigma_{F_{\rm beam}}^2}{F_{\rm beam}^2}\right],
\end{equation}
where \begin{equation}
  \sigma_{\log E_{\gamma}}^2=\sigma_a^2 + \left(\sigma_b\log\frac{E_{p,i}}{300~{\rm keV}}\right)^2 + \left(\frac{b}{\ln 10}\frac{\sigma_{E_{p,i}}}{E_{p,i}}\right)^2 + \sigma_{\rm int}^2.
\end{equation}

\begin{figure}
  \centering
 \includegraphics[width=12 cm]{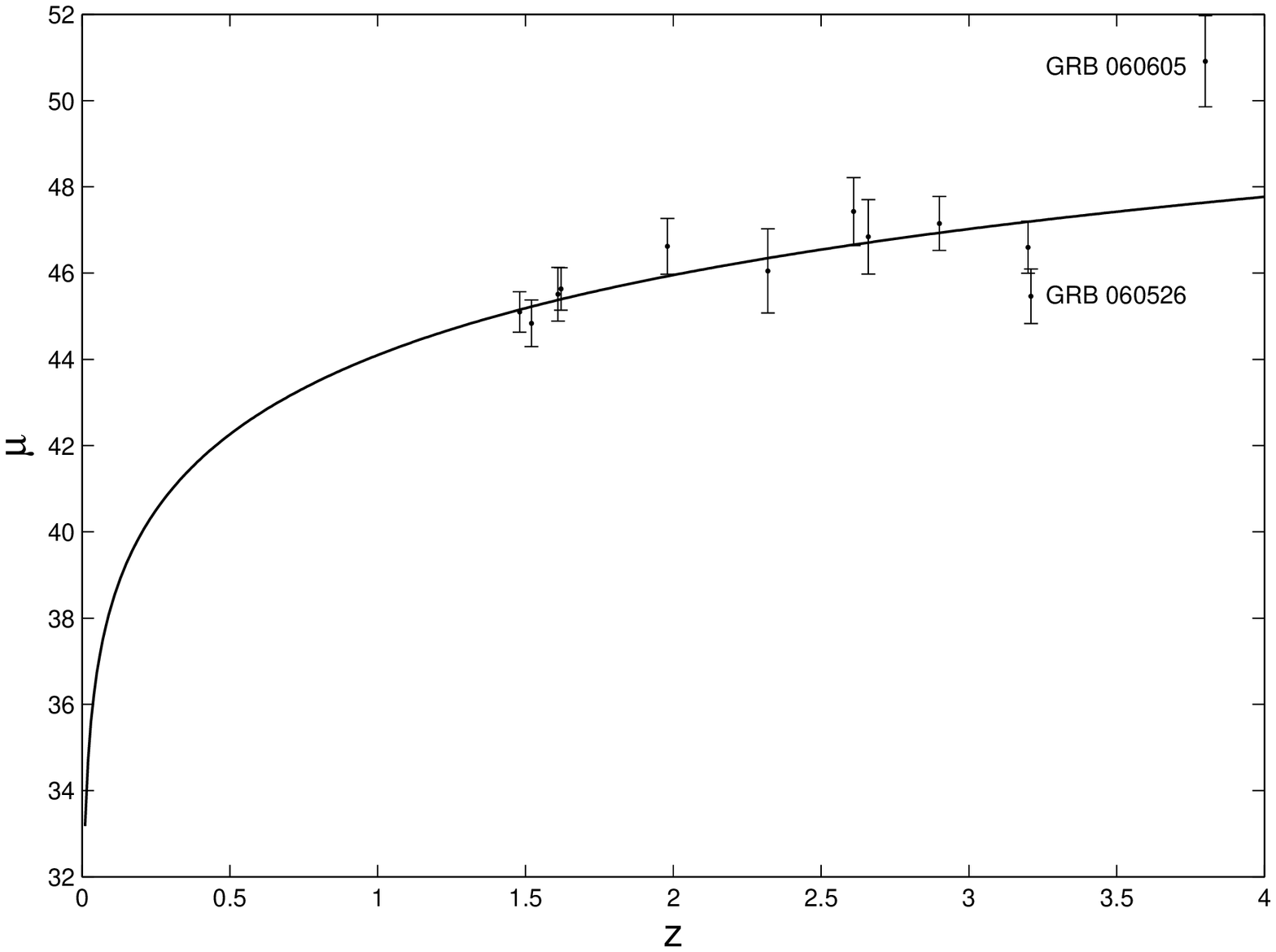}
 \caption{\small{The Hubble diagram of 12 long GRBs calibrated using the $E_p-E_{\gamma}$ relation. The black curve is the best-fit result to $\Lambda$CDM model. The best-fit parameter is $\Omega_M=0.302\pm 0.142$.}}\label{fig:hubble}
\end{figure}

The distance moduli of 12 high-$z$ GRBs and their $1\sigma$ uncertainties calibrated through the $E_p-E_{\gamma}$ relation are listed in Table \ref{tab:dis_moduli}.
\begin{table}
\centering
\caption{\small{The distance moduli of 12 high-$z$ GRBs calibrated through the $E_p-E_{\gamma}$ relation.}}\label{tab:dis_moduli}
\begin{tabular}{llll}
  \hline\hline
  GRB   & $z$ & $\mu$ & $\sigma_{\mu}$  \\
  \hline
  010222	&	1.48	&	45.0990	&	0.4684	\\
  030328	&	1.52	&	44.8357	&	0.5404	\\
  990123	&	1.61	&	45.5094	&	0.6244	\\
  990510	&	1.62	&	45.6337	&	0.4921	\\
  030226	&	1.98	&	46.6219	&	0.6476	\\
  021004	&	2.32	&	46.0496	&	0.9760	\\
  050820A	&	2.61	&	47.4273	&	0.7877	\\
  030429	&	2.66	&	46.8414	&	0.8608	\\
  050401	&	2.9	    &	47.1522	&	0.6272	\\
  020124	&	3.2	    &	46.5993	&	0.6032	\\
  060526	&	3.21	&	45.4623	&	0.6342	\\
  060605	&	3.8	    &	50.9127	&	1.0548	\\
  \hline
\end{tabular}
\end{table}
We also plot the 12 high-$z$ GRBs in $z-\mu$ plane in Figure \ref{fig:hubble}, where the black curve is the best-fit result to $\Lambda$CDM model.
The fit of 12 high-$z$ GRBs to the $\Lambda$CDM model gives $\Omega_M=0.302\pm 0.142$, well consistent with the Planck 2015 results \citep{Ade:2015xua}. From Figure \ref{fig:hubble}, we can see that the distance of GRB 060605 is much overestimated. This is because the $E_p-E_{\gamma}$ relation overestimates the energy of GRB 060605 (see also the $E_p-E_{\gamma}$ plot in Figure \ref{fig:correlation}, where the red star represents this burst). On the contrary, the distance of 060526 is underestimated because the $E_p-E_{\gamma}$ relation underestimates its energy. The rest 10 GRBs are consistent with the $\Lambda$CDM model within $1\sigma$ uncertainties.

For comparison, we also calibrate GRBs through the $E_p-E_{\rm iso}$ relation (the so called Amati relation). In this case, 40 low-$z$ GRBs and 61 high-$z$ GRBs are available. The constraint of 61 high-$z$ GRBs on the $\Lambda$CDM model gives $\Omega_M=0.805\pm 0.144$, which is much larger than the Planck 2015 results. The reason for this can be easily understood. From the $E_p-E_{\rm iso}$ plot in Figure \ref{fig:correlation}, we can see that high-$z$ GRBs have in average larger isotropic equivalent energy than low-$z$ GRBs at the same $E_p$ value. Therefore, when extrapolating the Amati relation from low-$z$ GRBs to high-$z$ GRBs, the energy (so the distance) of most high-$z$ GRBs is underestimated. The underestimation of distance further leads to the overestimation of $\Omega_M$. For this reason, we can predict that GRBs calibrated through the rest four luminosity correlations ($\tau_{\rm lag}-L$, $V-L$, $E_p-L$ and $\tau_{\rm RT}-L$) may also overestimate the value of $\Omega_M$.

\section{Summary}\label{sec:conclusions}

In this paper, we checked the possible redshift dependence of six luminosity correlation in long GRBs. We divided GRBs into low-$z$ and high-$z$ subsamples according to their redshift smaller or larger than 1.4. The slope and intercept parameters of six luminosity correlations are derived by maximizing the D'Agostini's likelihood. For all the six luminosity correlations, high-$z$ GRBs seem to have larger intercept, but smaller absolute slope than low-$z$ GRBs. It was shown that the intrinsic scatter of $V-L$ relation is to large to make a convincing conclusion. The $E_p-E_{\gamma}$ relation has the smallest intrinsic scatter among the six, although the number of available GRBs is small. Most importantly, the $E_p-E_{\gamma}$ relation shows weak redshift dependence. Strong evidence ($>3\sigma$) for the redshift evolution was found in the rest four correlations. Similar features can be seen when we use Reichart's likelihood instead of D'Agostini's, although the statistical significance is lower. We calibrated high-$z$ GRBs using the $E_p-E_{\gamma}$ relation in a model independent way and reconstruct the Hubble diagram. The constraint of high-$z$ GRBs on the $\Lambda$CDM model gives matter density $\Omega_M=0.302\pm 0.142$, which is well consistent with the Planck 2015 results, although the error bar is large. Calibrating GRBs using the Amati relation, as was done by \cite{Liu:2014}, in some cases may overestimate $\Omega_M$. One of the disadvantage in using the $E_p-E_{\gamma}$ relation, of course, is that only a small number of GRBs are available since most GRBs have no measurement of jet opening angle. We hope that the future observation will enlarge the GRB sample so as to improve the statistical significance.

\section*{Acknowledgements}
We are grateful to Y. Sang, P. Wang, S. Wang and D. Zhao for useful discussions. We thanks the anonymous referee for his/her useful comments and suggestions. We also appreciate G. Allodi for making the FMINUIT publicly available. This work has been funded by the National Natural Science Fund of China under grants Nos. 11375203, 11305181.

\label{lastpage}

\end{document}